# Towards Understanding the Impact of Real-Time AI-Powered Educational Dashboards (RAED) on Providing Guidance to Instructors

Ajay Kulkarni
George Mason University
Fairfax, VA
akulkar8@gmu.edu

Michael Eagle
George Mason University
Fairfax, VA
meagle@gmu.edu

## ABSTRACT

The objectives of this ongoing research are to build RealTime AI-Powered Educational Dashboard (RAED) as a decision support tool for instructors, and to measure its impact on them while making decisions. Current developments in AI can be combined with the educational dashboards to make them AI-Powered. Thus, AI can help in providing recommendations based on the students' performances. AIPowered educational dashboards can also assist instructors in tracking real-time student activities. In this ongoing research, our aim is to develop the AI component as well as improve the existing design component of the RAED. Further, we will conduct experiments to study its impact on instructors, and understand how much they trust RAED to guide them while making decisions. This paper elaborates on the ongoing research and future direction.

## Keywords

Decision support tool, Educational dashboard, Interactive visualizations, Impact, Unsupervised learning, Recommendations

## 1. INTRODUCTION

A dashboard is a collection of wisely selected visualizations that assists in understanding raw information stored in databases, which helps human cognition [6]. A dashboard can be viewed as a container of indicators [13], but Bronus et al. provided the most accurate definition of the dashboard. Bronus et al. defined the dashboard as "an easy to read, often single page, real-time user interface, showing a graphical presentation of the current status (snapshot) and historical trends of an organization key performance indicators (KPIs) to enable instantaneous and informed decisions to be made at a glance" [5]. This type of visual displays are critical in sense-making as humans are able to process large amounts of data if presented in a meaningful way [17]. The use of learning analytics tools and visualizations have the potential to provide effective support to instructors by helping them to keep students engaged and achieve learning objectives [15]. Yoo et al. [21] conducted a review of educational dashboards in which they underline the usefulness by mentioning dashboards present the results of the educational data mining process and help teachers to monitor and understand their student's learning patterns. We can apply the same principle to the data collected from a student's quiz questions. The responses received from the quiz can be used for understanding conceptual and meta-cognitive knowledge components [4]. It has also been noted that very few of the deployed learning dashboards addressed the actual impact on instructors [20]. Thus, we see a need for a Real-Time AIPowered Educational Dashboard (RAED) that is designed for assisting instructors. There are two main objectives of the proposed research.

**Objective 1:** Build a RAED, which will act as a decision support tool for instructors.

**Objective 2:** Measure the impact of the RAED on instructors and understand their trust in using the RAED while making decisions.

The proposed dashboard consists of two components - the visualization component and the AI component. The visualization component will present an entire classroom's actions in real-time on the dashboard. This will help instructors to answer two questions: (i) where are most of the students struggling? and (ii) on which questions are most of the students using hints? Answers to these questions can be useful for providing further explanations of certain concepts immediately after the quiz. The AI component of the dashboard will perform unsupervised learning on the collected responses. It will produce clusters of students and also generate recommendations based on the results, which will be displayed on the dashboard. These recommendations made by the AI component will also be included in the visualizations of the dashboard. These visuals will facilitate the instructors' decision-making process. For instance, an instructor may decide to give additional questions or teach a particular concept again after getting recommendations. Therefore, the current research will also be useful for understanding the usability, impact, and trust in the fusion of visualizations and AI in real-time.

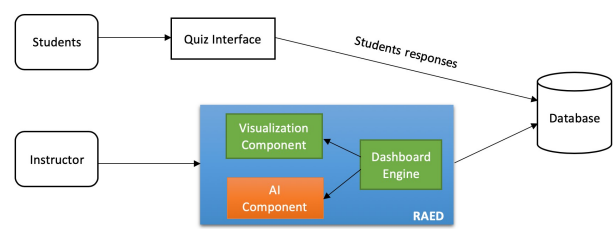

**Figure 1: Proposed architecture of the RAED**

## 2. RELATED WORK

There is a growing interest in the design and development of real-time systems, such as [14] and [19], that can provide actionable teaching analytics in real-time for decision making. These real-time systems are also beneficial from students' perspective because it gives more time to teachers to provide one-on-one support to students [11]. A user-centric teacher dashboard has been developed by Aleven et al. [2] for understanding interaction data and analytics from intelligent tutoring systems (ITS). Aleven et al. [2] noted that a dashboard could be useful to teachers for helping the class while teaching, and for preparation for the next classes. Diana et al. [7] displayed real-time analytics of interactive programming assignments on an instructor dashboard. From the results, Diana et al. [7] concluded that student outcomes could be accurately predicted from the student's program states. In addition to that, for helping more students in a classroom, Diana et al. [8] also used the machine learning model along with approach maps for identifying and grouping students who need similar help in real-time. Holstein et al. [12] developed the Luna dashboard by collecting data from interpretation sessions and affinity diagramming from middle-school teachers. The goal was to understand the dashboard's usability from the teacher's aspect, as well as its effect on students learning. In a recent paper [10], a wearable classroom orchestral tool for K-12 teachers was tested by Holstein et al. The classroom was represented as a dashboard. In that research, mixed-reality smart glasses were connected to ITSs for understanding real-time student learning and behavior within the ITSs. A framework consisting of five dimensions (Target, Attention, Social visibility, Presence over time, and Interoperation) has also been proposed for the design and analysis of teaching augmentation in [3].

## 3. PROPOSED CONTRIBUTIONS

This section presents the architecture and design of our proposed RAED. It further describes the features of the RAED and discusses its desirable properties, such as portability and explainability. It also includes information on the current state of our RAED development.

### 3.1 Architecture and design

The architecture of RAED is shown in Figure 1. Students will get a quiz interface on which they will see the questions and respond to them. The responses will get stored in a database, which can be queried by the dashboard engine. The dashboard engine will be responsible for data preprocessing and data cleaning. The resulting clean data will then be given as input to the visualization and AI components. The visualization component will produce visualizations on the dashboard, and the AI component will perform clustering of the data in real-time. The results from the AI component will be visualized and interpreted in terms of recommendations. Currently, we have developed a quiz interface [1] for experimental purposes, using R and Shiny. This interface stores results on a Google sheet, and currently this Google sheet acts as our database. The dashboard engine is connected to the Google sheet, which queries data every 6 seconds. Thus, the dashboard is refreshed every 6 seconds.

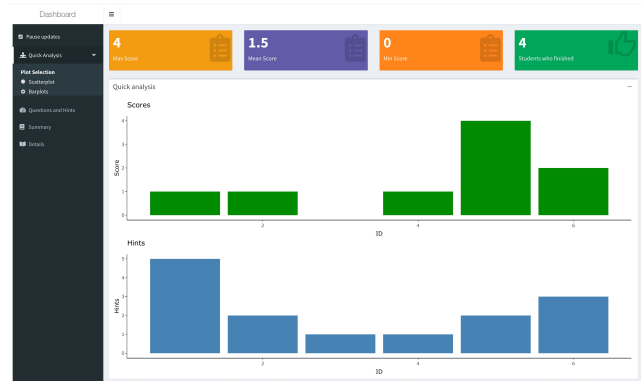

**Figure 2: Design component of the RAED**

In this on-going research, we have implemented a design component of the dashboard (shown in Figure 2), which displays real-time visualizations [2]. The essential characteristics of the dashboards noted by Few [9] are taken into consideration while designing our dashboard. We also will be following four elements of the learning analytics process model [20] as a foundation for the conceptual design. These four elements are awareness, reflection, sensemaking, and impact. At the current state, our design includes the first three. Awareness refers to the data, which can be visualized or represented in tables as it streams. Reflection focuses on mirroring teaching practices, and sensemaking can deal with the understanding at-risk students [20].

### 3.2 Features of the dashboard

We propose five unique features of the RAED. We have implemented a majority of the features, and details are as follows.

**1) Interactive visualizations** – The visualizations generated on the dashboard are fully interactive (shown in Figure 3) and can be downloaded in Portable Network Graphics (PNG) format. Instructors can interact with them by zooming in, zooming out, selecting different components of the visualizations, etc. These visualizations can also provide meaningful information if the cursor hovers over them. Currently, our dashboard visualizations include scatter plots, bar plots, and histograms. The scatter plot and bar plots are used for understanding the quiz score and number of hints requested by students. Histograms are used to understand the score distribution of the class. Another essential role of the RAED is enhancing the perception of instructors as they can decide what to focus on.

---

[1] https://tinyurl.com/qnp46y9

[2] https://tinyurl.com/yx3pht5e

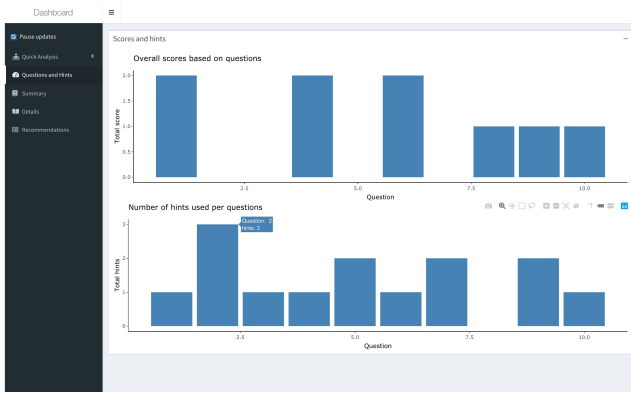

**Figure 3: Interactive visualizations of the RAED**

**Figure 4: Dynamic tables on the RAED**

**2) Dynamic tables –** One of the unique features of the RAED is its dynamic tables. We have provided a summary of the class and scorecard in the form of tables. These tables get updated in real-time, and instructors can search as well as sort the tables. The RAED also provides functionality to download these tables into CSV format for further analysis (shown in Figure 4).

**3) Portability –** The dashboard is designed in R and deployed on the Shiny server. This dashboard is portable and can be connected to any database or tool.

**4) Real-time –** The dashboard provides updates of the data every 6 seconds, which can help to capture student's real-time interaction during the quiz. Further, we provide the additional feature of pausing and resuming real-time streaming. This feature can be especially useful when analyzing the dashboard.

**5) AI and explainability –** This feature is currently under development. We plan to employ explainable AI on the dashboard. It will help instructors to understand how AI provides results to the dashboard, i.e., how it chooses the number of clusters and how it produces recommendations.

## 4. FUTURE DIRECTION

The future direction of this research is to develop a prototype of the RAED, test it in classrooms, and then conduct surveys to measure its impact and trust. Future research will be held in the following four phases.

- **Phase 1 (AI component):** Currently, we can store student ids, names of the course topics, responses, scores, whether hints are used, and what is the total number of requested hints. We will be using this information for clustering students and generating recommendations for them. The process of clustering can be useful for focusing attention on students with similar characteristics and learning rates. This information can help instructors to form support groups within the class and to provide personalized guidance to particular students. For example, students from the highperformance group can be paired with students in the low-performance group, which can help to improve performance.

  In the first step, similar students will be identified by performing clustering on the data. The goal of this step is to identify three clusters (high performance, average performance, and low performance). It is essential to visualize the process of clustering for implementing explainable AI. Thus, the implementation of Agglomerative Hierarchical Clustering makes a suitable choice. Using this approach, clustering process will begin with points as individual clusters, and at each step, the similar points will be merged into a larger cluster [18]. This entire process of clustering can be visualised by plotting a dendrogram, which fulfills our goal of explainability. The other advantage of using Agglomerative Hierarchical Clustering is that it provides good results when given small datasets as input [1], as is the expected number of students in a class. In the next step, information of students from these clusters will be obtained, and a list of concepts that students need to improve will be derived from the responses. It will also provide suggestions on pairing students during in-class activities. This information will act as recommendations to instructors and can also help to understand conceptual as well as metacognitive knowledge components of the class.

- **Phase 2 (Design):** We will focus on the design aspect of the RAED using the iterative design process. In this step, the prototype will be shown, and functionalities will be explained to the instructors for getting their insights on RAED. Surveys will be provided to the instructors for evaluations and to know their additional needs. Questions in the survey will be based on the questionnaire created by Park et al. [16].

  The results will help us to get inputs on information usefulness, visual effectiveness, appropriateness of visual representation, user-friendliness, and understanding of the information. Changes will be made in the design after analyzing responses from the survey. In the next iteration, RAED will be shown again to the instructors, and their feedback will be requested. In this way, at the end of this phase, the prototype will be ready for testing.

- **Phase 3 (Testing):** In this phase, the prototype will be tested in classrooms. The quiz interface will be provided to students, and RAED will be made available to instructors.

This phase will help us understand the technical problems that may occur, such as server issues. Improvements will be made as necessary.

- **Phase 4 (Survey):** In this final phase, surveys will be provided to instructors to understand their changes in behavior, the achievement of the objective, trust in the system, effect on motivation and decision making due to the RAED. The responses will help us to measure the impact of the RAED on the instructor's decision making. It will also give us insights about the trust instructors have in the RAED.

## 5. ACKNOWLEDGMENTS

The authors would like to thank DataLab at George Mason University for their support. The authors also would like to thank Dr. Olga Gkountouna for useful feedback on this work.